\documentstyle[12pt,psfig]{article}
\begin{document}
\setlength{\unitlength}{1mm}

{\hfill   November 1998 }

{\hfill   SPIN-1998/10}

{\hfill    hep-th/9812056} \vspace*{2cm} \\
\begin{center}
{\Large\bf Conformal description of horizon's states}
\end{center}
\begin{center} 
Sergey N.~Solodukhin 
\end{center}
\begin{center}
 Spinoza Institute,
University of Utrecht, 
Leuvenlaan 4, 
3584 CE Utrecht, \\
the Netherlands
\end{center}
\begin{center}
e-mail: S.Solodukhin@phys.uu.nl
\end{center}
\vspace*{1cm}
\begin{abstract}
\noindent
The existence of black hole horizon is considered as a boundary
condition to be imposed on the fluctuating metrics. The coordinate invariant
form of the condition for class of spherically symmetric metrics is formulated.
The diffeomorphisms preserving   this condition act in (arbitrary small)
vicinity of the horizon and form the group of conformal transformations
of two-dimensional space
($r-t$ sector of the total space-time). The corresponding algebra recovered
at the horizon is one copy of the Virasoro algebra. For general
relativity in $d$ dimensions we find an effective two-dimensional theory which governs the conformal 
dynamics at the horizon universally  for any $d\geq 3$. The corresponding Virasoro algebra has 
central charge $c$ proportional to the Bekenstein-Hawking entropy. Identifying the zero-mode 
configuration we calculate $L_0$. The counting of states of this horizon's conformal field theory
by means of Cardy's formula is in complete agreement with the Bekenstein-Hawking
expression for the entropy of black hole in $d$ dimensions.
\end{abstract}
\newpage
\section{Introduction}
\setcounter{equation}0
Since the remarkable  discovery \cite{BH} that a black hole has entropy proportional
to the area of horizon
\begin{equation}
S_{BH}={A_h\over 4G}
\label{1}
\end{equation}
it remains a mystery as what states are counted by this formula.
A number of approaches was proposed  \cite{2} to answer this question
within a conventional field theory. However, reproducing correctly the proportionality
to the area  most of these approaches normally lead to divergent expression
for the entropy.

On the other hand, there is a number of indications that two-dimensional
conformal symmetry may provide us with relevant description of black hole's
states. The Hilbert space of a conformal field theory realizes a representation of the 
(quantum) Virasoro
algebra
\begin{equation}
[L_n,L_m]=(n-m)L_{n+m} +{c\over 12}n(n^2-1)\delta_{n+m,0}
\label{2}
\end{equation}
with infinite set of generators $L_n$ and central charge $c$. The number of states  in the conformal 
field theory at the level $L_0$
grows exponentially and the corresponding entropy is given by
\begin{equation}
S_{conf}=2\pi\sqrt{cL_0\over 6}~~.
\label{3}
\end{equation}
In the current literature, this formula is known  as Cardy's formula.
That entropy (\ref{3}) of an appropriately defined conformal field theory may fit
the expression (\ref{1}) was demonstrated in number of examples found within 
string theory \cite{3}. However, the relevant conformal field theory lives in
flat space-time and additional arguments should be given to relate its states to the
ones living in the black hole phase.

Another example which is inspiring for the present consideration is three-dimensional 
BTZ black hole \cite{BTZ}.   The feature of general relativity in three dimensions is
that it can be re-formulated as $SL(2,R)\times SL(2,R)$ Chern-Simons theory with
only dynamical degrees of freedom living on the boundary. It was argued by Carlip \cite{Carlip}
some time ago that  states of the  boundary theory realizing a representation of (two copies of)
Virasoro algebra are responsible for the entropy of BTZ black hole.
An alternative elegant calculation was proposed by Strominger \cite{Strominger}.
He uses the fact proven in \cite{Brown-Henneaux} that imposing boundary condition
that 3d metric is asymptotically  AdS$_3$ there is group of diffeomorphisms
preserving this condition. This group is generated by two copies of the Virasoro algebra
with central charge $c={3l\over 2G}$ ($G$ is Newton's constant and $l$ is AdS$_3$ radius).
The relevant conformal theory is the Liouville theory \cite{Liouville} living on the
(two-dimensional) boundary
and described by the action
\begin{equation}
W_L=\int d^2z\sqrt{-\gamma} \left( {1\over 2} (\nabla \phi )^2+Q\phi R+\lambda e^{-2\phi}\right)~~,
\label{4}
\end{equation}
where $Q$ is related to the central charge $c={3l\over 2G}$ as  $c=48\pi Q^2$.
The BTZ metric obeys the boundary condition and the black hole, thus, is in the Hilbert space
of the conformal field theory. The counting of the degeneracy using the formula
(\ref{3}) then exactly reproduces the Bekenstein-Hawking result.
This calculation, however, 
uses features specific for the three-dimensional gravity and
it is not seen how it  can  be extended
to higher dimensions.

A goal of the present paper is to follow a similar line of reasoning but not restricting
to specific features of three dimensions. The key idea is to formulate a boundary condition
describing metrics with black hole horizon. Indeed, the existence of the horizon is a
requirement which essentially restricts class of possible metrics.
The coordinate invariant form of this condition is formulated in the next section 
and is appropriate to describe
both dynamical and static black holes. Remarkably, we find  diffeomorphisms
that preserve this condition. They act in (arbitrary small) vicinity of the horizon
and form the infinite-dimensional group of conformal
transformations in two dimensions. 
The corresponding algebra recovered at the horizon is one copy of the Virasoro algebra. The physics
at the horizon is, thus,  conformal that could be anticipated since it is known that 
fields becomes effectively massless at the horizon.
For general relativity the conformal 
dynamics at the horizon is governed by an effective two-dimensional field theory. It is
constructed in Section 3 and is shown in Section 5 to be universal
theory at the horizon of black hole in any space-time dimensions. The 
counting states of this horizon's conformal field theory by Cardy's formula (\ref{3}) 
exactly reproduces the result (\ref{1}). Note, that
in our approach the states responsible for the entropy (\ref{1}) are states of the horizon itself. This
differs from the picture present in \cite{Strominger} where the relevant states live at infinity.

\section{ Horizon boundary condition and 2d conformal group}
\setcounter{equation}0
In a theory of quantum gravity dealing with fluctuating space-time geometry
one should be able to formulate conditions which fix the class of possible metrics.
One of the conditions one usually considers is the behavior of metric at
infinity. The space-time is then supposed to be asymptotically flat or asymptotically 
(Anti-)  Sitter dependent on the physical situation. However, the fixing  
of the asymptotic behavior not completely specifies the topology
of the space-time. This should be considered as an additional requirement restricting
the class of metrics under consideration. The presence of the black hole horizon
is such topological feature which should be traced in the conditions which one should
impose  on
the space-time metrics. There are different definitions of horizon, some of them, in
particular, require knowledge of global (all-time) behavior of space-time.
More appropriate for our goals is the notion of $\it apparent$ horizon which can
be defined locally as the boundary of trapped region \cite{Wald}.

Consider four-dimensional spherically symmetric metric
of the general form
\begin{equation}
ds^2=\gamma_{ab}(x^0,x^1)dx^adx^b+r^2(x^0,x^1)(d\theta^2+\sin^2\theta d\phi^2)
\label{2.1}
\end{equation}
where $\gamma_{ab}(x^0,x^1)$ can be considered as metric on effective 2d space-time
$M^2$ with coordinates $x^0,x^1$; the radius $r(x^0,x^1)$ is then scalar function on
$M^2$. For this class of metrics the apparent horizon can be defined \cite{Russo}
as a curve $\cal H$
on $M^2$ such that the gradient of the radius $r(x^0,x^1)$
\begin{equation}
\gamma^{ab}\nabla_ar\nabla_br|_{\cal H}=0
\label{2.2}
\end{equation}
vanishes along $\cal H$. This condition is invariant under conformal transformation
$\gamma_{ab}\rightarrow e^{2\rho}\gamma_{ab}$ where $\rho$ is a regular on $\cal H$
function. Therefore, the 2d metric $\gamma_{ab}$ at the horizon
is determined by the condition (\ref{2.2}) only up to a (regular) conformal factor.

It is convenient to use conformal coordinates $x_+,x_-$ in which the 2d part of the metric
(\ref{2.1}) takes the form $\gamma_{ab}(z^0,z^1)dz^adz^b=-e^{2\sigma (x_+,x_-)}dx_+dx_-$.
Then locally we may choose coordinates $x_+,x_-$ in such a way that the equation for the curve $\cal H$
becomes $x_-=0$. Assuming that the function $\sigma (x^0,x^1)$ is regular at $x_-=0$,
we find that the equation (\ref{2.2}) reads
\begin{equation}
\partial_+r(x_+,x_-)|_{x_-=0}=0~~.
\label{2.4}
\end{equation}

Note, that the condition (\ref{2.2}) is appropriate to describe dynamical 
black hole. The horizon, in general,  may consist on different components. Then the condition
(\ref{2.2}) locally defines  each component. For example, in static case the horizon has two
intersecting components ${\cal H}_+$ ($x_-=0,~x_+>0$) and ${\cal H}_-$ ($x_+=0,~x_-<0$). 
The later is defined as
\begin{equation}
\partial_-r(x_+,x_-)|_{x_+=0}=0~~.
\label{2.5}
\end{equation}
In what follows we consider only one component of the horizon defined by (\ref{2.4}).

Expanding the function $r(x_+,x_-)$ near $x_+=0$ we find 
\begin{equation}
r(x_+,x_-)=r_h+\lambda (x_+)x_- +O(x^2_-)
\label{2.6}
\end{equation}
that is consistent with the condition (\ref{2.4}); $\lambda$ is  an arbitrary function
of $x_+$, $r_h$ is constant (radius of horizon). It follows from (\ref{2.6})
that
\begin{equation}
\partial_+ r=\partial_+\lambda (x_+)x_-+O(x_-^2)
\label{2.7}
\end{equation}
in vicinity of $\cal H$. Now it is not difficult to find group of diffeomorphisms which preserve the
condition (\ref{2.2}), (\ref{2.4}), (\ref{2.6}). Indeed, consider vector 
$\xi_+=(\xi^+_+=g(x_+), \xi^-_+=0,0,0)$. Then using (\ref{2.7}) we have  
$$
{\cal L}_{\xi_+}(\partial_+ r)=\partial_+(\xi_+^+ \partial_+ r)=\partial_+(g(x_+)\partial_+
\lambda (x_+))
x_-+O(x^2_-)~~
$$
for the Lie derivative along the vector $\xi_+$.
Thus, the condition (\ref{2.4}),  (\ref{2.6}), (\ref{2.7}) is preserved under diffeomorphisms generated by $\xi_+$
provided that the function
$\lambda (x_-)$ changes as follows
$$
{\cal L}_{\xi_+}\lambda =g(x_+)\partial_+\lambda (x_+)~~.
$$
This is standard transformation law for a scalar under diffeomorphisms. Note, 
that $\lambda (x_+)$ in (\ref{2.6})
is indeed  scalar since it can be represented in the covariant form as
$$
n^\mu \partial_\mu|_{\cal H} ~r=\lambda (x)~~,
$$
where $n_\mu$ ($n^2_\mu=0$) is normal to $\cal H$ and $\lambda (x)$ is function along $\cal H$.

Diffeomorphisms generated by vector $\xi_+$ are  tangential
to $\cal H$ and thus preserve $\cal H$. In the case of horizon with two
bifurcating components ${\cal H}_+$ and ${\cal H}_-$ vector $\xi_+$ generates symmetry of ${\cal H}_+$ while
the component ${\cal H}_-$ is invariant under diffeomorphisms generated by vector
$\xi_-=(\xi^+_-=0,\xi^-_-=f(x_-),0,0)$. Note, that both vectors $\xi_+$ and $\xi_-$ satisfy the equation 
$$
\nabla_a\xi_b+\nabla_b\xi_a={1\over 2}\gamma_{ab}\nabla_c\xi^c
$$
for two-dimensional
conformal Killing vectors
and thus generate the infinite-dimensional group of conformal  transformations of the space $M^2$.
The corresponding generators  $l^\pm_n=e^{\imath n x_\pm}\partial_\pm$ form two copies of the 
Virasoro
algebra
\begin{equation}
[l^\pm_n,l^\pm_m]=\imath (m-n)l^\pm_{n+m}~~
\label{2.8}
\end{equation}
with respect to the Lie bracket $[\xi_1,\xi_2]=(\xi_1\xi_2'-\xi_2\xi_1')\partial_x$.
However, near each component (${\cal H}_+$ or ${\cal H}_-$)
of $\cal H$ there is only one copy of the Virasoro algebra which leaves the horizon invariant.
As in the case of 2d space with boundary \cite{Cardy} 
the presence of horizon breaks one of the conformal 
symmetries.
Note, that the Virasoro algebra (\ref{2.8}) is algebra of diffeomorphisms of non-compact space
$R^1$. In particular, this means that $n$ and $m$ in (\ref{2.8}) are arbitrary  numbers and not
necessarily  integers. Usually, one considers the compact version of the Virasoro
algebra which is algebra of diffeomorphisms of circle $S^1$. For the further purposes we need
the compact version of the algebra. Therefore, we consider an arbitrary large interval $L$ of $R^1$ and impose 
periodic boundary conditions. At the end we take $L$  to be infinite.

It should be noted that we did not use so far any
gravitational field equations and defined a general class of (spheri-symmetric) metrics
with black hole horizon.
We can also see  from the present analysis that the condition (\ref{2.2})
defying the horizon $\cal H$ is essentially a condition on the function
$r(x^0,x^1)$ in the 4d metric (\ref{2.1}) while the form of the 2d metric $\gamma_{ab}(x^0,x^1)$
remains undetermined. Another indication of this is the fact that the condition
(\ref{2.2}) does not change  under the conformal transformation of the 2d metric.
Thus, the condition (\ref{2.2}) defines a class of  metrics (\ref{2.1}) modulo
this conformal transformation.
Therefore, the only gravitational dynamics of the fluctuating (of-shell) 4d metric
arising on the horizon is the dynamics of the radial function $r(x^0,x^1)$ 
while for the 2d part one can take any metric from the same conformal class.
In what follows we consider the static case and 
choose the representative  metric on 2d space-time $M^2$ in the form
\begin{equation}
ds^2=-g(x)dt^2+{dx^2\over g(x)}~~
\label{2.9}
\end{equation}
with the function $g(x)$  vanishing at $x=x_h$, where $x_h$ is location of the 
horizon $\cal H$ in the 
Schwarzschild coordinates ($t,x$). In assumption that we deal with a 
non-extreme black hole we have that
\begin{equation}
g(x)= {2\over \beta_H}(x-x_h)+O((x-x_h)^2)~~,
\label{2.10}
\end{equation}
where $\beta_H$ is constant related to the surface gravity of the horizon.

\section{ Bekenstein-Hawking entropy as central charge of the Virasoro algebra}
\setcounter{equation}0
It follows from the consideration of the previous Section that any theory of quantum gravity describing black hole should provide us with a realization of the Virasoro algebra in the region close to horizon.
In this Section we demonstrate how it works for  general relativity.

We start with four-dimensional Einstein-Hilbert action
\begin{equation}
W_{EH}=-{1\over 16\pi G} \int_{M^4}d^4z\sqrt{-g}R_{(4)}
\label{3.1}
\end{equation}
and consider it on the class of spherically symmetric metrics
(\ref{2.1}). We arrive at an effective two-dimensional theory
described by the action
\begin{equation}
W=-\int_{M^2}d^2x\sqrt{-\gamma}\left( {1\over 2}(\nabla \Phi )^2+{1\over 4}\Phi^2 R +{1\over 2G} \right)~~,
\label{3.2}
\end{equation}
where $\Phi=r G^{-1/2}$ and $R$ is 2d scalar curvature. This action takes the form of dilaton gravity (the radius $r$ playing
the role of dilaton field) in two
dimensions and can be transformed to the form similar to that of
the Liouville theory (\ref{4})
\begin{equation}
W=-\int_{M^2}\left({1\over 2}(\nabla\phi )^2+{1\over 4}q\Phi_h \phi R+U(\phi )\right)
\label{3.3}
\end{equation}
by applying the transformation \cite{RT}
\begin{equation}
\gamma_{ab}=({\phi_h\over \phi })^{1\over 2} e^{{2\over q\Phi_h}\phi} \bar{\gamma}_{ab}~~,~~
\phi={1\over q}{\Phi^2\over \Phi_h}~~,
\label{3.4}
\end{equation}
where $\Phi_h=r_hG^{-1/2}$ is the classical value of the field $\Phi$ on the horizon, i.e.
the horizon radius $r_h$ measured
in the Planck units. The classical value of the field $\phi$ is respectively $\phi_h=q^{-1}\Phi_h$.
Since we are interested in the region very close to horizon, where the 2d metric $\gamma_{ab}$
is determined up to conformal factor, we obtain an equivalent system provided that the
conformal transformation (\ref{3.4}) is regular at the horizon.
The action (\ref{3.3}) depends on an arbitrary constant $q$. So does the central charge which we
will calculate in a moment. However, the final result (the statistical entropy counted in the next
section) is independent of $q$.  The potential $U(\phi )$ in (\ref{3.3}) is
$$
U(\phi )={1\over 2G}({\phi_h\over \phi })^{1\over 2} e^{{2\over q \Phi_h}\phi}
$$
but its form is not important for our consideration.

Varying the action (\ref{3.3}) with respect to the dilaton $\phi$ and metric $\bar{\gamma}_{ab}$
we obtain the equation of motion for $\phi$
\begin{equation}
\Box \phi ={1\over 4}q\Phi_hR+U'(\phi)
\label{3.5}
\end{equation}
as well as constraints
\begin{equation}
T_{ab}\equiv {1\over 2}\partial_a\phi\partial_b\phi-{1\over 4}\bar{\gamma}_{ab}(\nabla\phi )^2
+{1\over 4}q\Phi_h (\bar{\gamma}_{ab}\Box \phi -\nabla_a\nabla_b\phi)-{1\over 2}\bar{\gamma}_{ab}U(\phi)=0~~.
\label{3.6}
\end{equation}
The theory of the scalar field $\phi$ described by the action (\ref{3.3}) is not conformal.
Indeed, we find that  the trace of (\ref{3.6})
\begin{equation}
\bar{\gamma}^{ab}T_{ab}={1\over 4}q\Phi_h\Box\phi -U(\phi )
\label{3.7}
\end{equation}
does not vanish. However, the theory becomes conformal being considered
in infinitely small vicinity of the horizon.
The 2d metric there takes the form (\ref{2.9})-(\ref{2.10}). Operating with this metric
it is convenient to use the coordinate
$$
z= \int^x{dx\over g(x)}={\beta_H\over 2}\ln (x-x_h)~~
$$
so that the vicinity of the horizon  ($(x-x_h)$ is small) looks as region of
infinite negative $z$. The metric function (\ref{2.10}) 
$$
g(z)=g_0e^{{2\over \beta_H}z}
$$
exponentially vanishes at the horizon. In the coordinates $(t,z)$ the equation (\ref{3.5})
reads
\begin{equation}
-\partial_t^2\phi+\partial_z^2\phi={1\over 4}q\Phi_h Rg(z)+g(z)U'(\phi )~~.
\label{3.8}
\end{equation}
Note, that the 2d scalar curvature $R$ may be non-zero at the horizon due to terms $\sim (x-x_h)^2$
present in the metric and neglected in (\ref{2.10}). We see that due to the exponentially
decaying factor $g(z)$ the r.h.s. of eq.(\ref{3.8}) exponentially vanishes for large negative $z$.
Therefore, in the region very close to horizon (infinite $z$) the r.h.s. of (\ref{3.8})
becomes negligible and we obtain the equation
\begin{equation}
\partial_t^2\phi-\partial_z^2\phi=0
\label{3.9}
\end{equation}
describing free  field propagating in flat space-time with coordinates
$(t,z)$. 
Expressing the constraints (\ref{3.6}) in terms of the coordinates
$(t,z)$ we  find that
\begin{eqnarray}
&&T_{00}={1\over 4}((\partial_t\phi )^2+(\partial_z\phi )^2)-{q\Phi_h\over 4}(\partial_z^2\phi-
{g'_x\over 2}\partial_z\phi ) +{1\over 2}g(z)U(\phi )~~,\nonumber \\
&&T_{0z}={1\over 2}\partial_t\phi\partial_z\phi -{q\Phi_h\over 4}(\partial_z\partial_t\phi-
{g'_x\over 2}\partial_t\phi ) ~~~, \nonumber \\
&&T_{zz}={1\over 4} ((\partial_t\phi )^2+(\partial_z\phi )^2)+{q\Phi_h\over 4}(-\partial_t^2\phi+
{g'_x\over 2}\partial_z\phi )-{1\over 2}g(z) U(\phi )~~.
\label{constraints}
\end{eqnarray}
The trace is
$$
-T_{00}+T_{zz}={1\over 4}q\Phi_h \left(-\partial_t^2\phi+\partial_z^2\phi \right)-g(z)U(\phi )~~.
$$
In the region of large $z$ this quantity vanishes on the equation
of motion (\ref{3.9}). This, in particular, guarantees that the Poisson algebra of constraints
(\ref{constraints}) closes and and in the region of infinite $z$ they form the Virasoro algebra.
We conclude that {\it the theory of the scalar field $\phi$  described by the action (\ref{3.3})
 is conformal being considered at the horizon (actually, in arbitrary small vicinity of the
horizon)}. The conformal transformations are generated by  charges
\begin{equation}
T[\xi]=\int_{-L/2}^{L/2}dz T_{++}\xi (z)~~,
\label{3.10}
\end{equation}
where 
\begin{eqnarray}
&&T_{++}=T_{00}+T_{0z} \nonumber \\
&&={1\over 4}(\partial_t\phi+\partial_z\phi )^2-{1\over 4}q\Phi_h\left( \partial_z (\partial_z+\partial_t)\phi-{1\over \beta_H}(\partial_z+\partial_t )\phi \right)~~.
\label{3.11}
\end{eqnarray}
A general solution of the eq.(\ref{3.9}) is sum of right- and left-moving
plane waves $\phi=\phi_+(t+z)+\phi_-(t-z)$. However, only the right-moving part contributes to 
$T_{++}$. It is worth noting that the present analysis can be done in terms of the original scalar field $\Phi$
(\ref{3.2}). Then in the limit of large $z$ 
the trace of the corresponding stress-energy tensor vanishes
under additional condition that 
$-(\partial_t\Phi)^2+(\partial_z\Phi )^2=0$ which indicates that one should consider only 
a part of modes. This is also consistent with our discussion in Section 2.
  
In the region of large $z$ we may use the translation invariance $z\rightarrow z+Z, Z=const$
in order to adjust $z$ to lie in the interval $-{L\over 2}\leq z\leq {L\over 2}$.
After all we take the limit of infinite $L$. Considering field $\phi$ on this interval we assume 
the periodic boundary condition to be imposed. The vector field $\xi (z)$ is also periodic,
$\xi(z+L)=\xi (z)$.

From (\ref{3.3}) we find the Poisson algebra
$$
\{\phi (z,t),\partial_t\phi (z',t)\}=\delta (z-z')
$$
and, as a consequence, we have
$$
\{(\partial_t+\partial_z)\phi(z,t), (\partial_t+\partial_z )\phi (z',t)\}=
\partial_{z}\delta (z-z')-\partial_{z'}\delta (z-z')~~.
$$
We now in a position to compute the Poisson algebra of the charges (\ref{3.10}). The result is
\begin{equation}
\{ T[\xi_1],T[\xi_2]\}=T[[\xi_1,\xi_2]]+({q\Phi_h\over 4})^2\int_{-L/2}^{L/2}C[\xi_1,\xi_2]dz~~,
\label{3.12}
\end{equation}
where $[\xi_1,\xi_2]=\xi_1\xi_2'-\xi_2\xi_1'$ and
$$
C[\xi_1,\xi_2]=(\xi_1'+\beta_H^{-1}\xi_1)(\xi_2'+\beta_H^{-1}\xi_2)'-
(\xi_1'+\beta_H^{-1}\xi_1)'(\xi_2'+\beta_H^{-1}\xi_2)
$$ 
is a deformation of the well known 
two-cycle $(\xi'_1\xi''_2-\xi_2'\xi_1'')$ of the algebra of 
$Diff(S^1)$.
That (\ref{3.12}) is identical to the  Virasoro algebra (\ref{2.8}) is easy to recognize by expanding 
$\xi (z)$ in Fourier series $\xi_n=e^{\imath {2\pi\over L}nz}$ and introducing the
Virasoro generators
\begin{equation}
L_n={L\over 2\pi}\int_{-L/2}^{L/2}dz e^{\imath {2\pi\over L}nz} T_{++}~~.
\label{3.13}
\end{equation}
They form the  algebra
\begin{equation}
\imath \{L_k,L_n \}=(k-n) L_{n+k}+ {c\over 12} k(k^2+({L\over 2\pi\beta_H})^2)\delta_{n+k,0}
\label{3.14}
\end{equation}
with the central charge $c=3\pi q^2\Phi^2_h$. Since the Bekenstein-Hawking entropy is
$S_{BH}=\pi\Phi^2_h$ we thus obtain that the central charge of the Virasoro algebra
(\ref{3.14})
\begin{equation}
c=3q^2S_{BH}
\label{3.15}
\end{equation}
is proportional to the Bekenstein-Hawking entropy.

\section{ Zero-mode configuration and the counting of the  states}
\setcounter{equation}0
In order to use Cardy's formula (\ref{3}) and count the number of states
we need to know value of $L_0$. It is typically determined by zero mode configuration
which is, in fact, classical configuration. In our case the classical configuration
(up to exponentially small terms $\sim O(e^{{2\pi\over \beta_H}})$) is just a
constant $\phi=\phi_h$. But $L_0$ vanishes for this configuration.
To resolve this problem note that a more general zero-mode configuration is
allowed to exist near the horizon
\begin{equation}
\phi_0=\alpha+Pz~~,
\label{4.1}
\end{equation}
which is obviously a solution of the field equations. This configuration infinitely grows 
close to horizon and should be excluded in the region  of infinite  $z$. 
But it may present if we consider our system in a box.
In order to make (\ref{4.1}) periodic with the period $L$ we first consider 
this function on the interval $0\leq z\leq L/2$ and then continue it to the interval
$-L/2\leq z\leq 0$ as $\phi_0(-z)=\phi_0 (z)$. Our condition for the function $\phi_0$ is that
it becomes $\phi_h$ at the right end of the interval
$$
\phi_0|_{z= {L\over 2}}=\phi_h~~.
$$
At the  point $z=0$ we impose  the condition
\begin{equation}
(\partial_z\phi_0+\beta^{-1}_H \phi_0 )|_{z=0}=0~~,
\label{cond}
\end{equation}
which means that  for the zero-mode the boundary term
$(n^a\partial_a+k)\phi$ (where $n^a$ is normal and $k$ is extrinsic curvature)
which appears in the (on-shell) gravitational action vanishes on the inner boundary $z=0$. 
Both conditions result in the following form of the configuration (\ref{4.1})
\begin{equation}
\phi_0=2\phi_h({z-\beta_H\over L-2\beta_H})
\label{4.2}
\end{equation}
in the interval $0\leq z\leq L/2$,
that gives us  $P=\partial_z \phi_0\simeq {2\phi_h\over L}$ for large $L$.
We see, in particular, that $P$ vanishes when $L$ becomes infinitely large
that is in accordance with our wish 
to have constant as the
classical configuration in the infinite region. 
Provided that we are happy with the imposed conditions the
value of $L_0$ is calculated as follows
\begin{equation}
L_0={L^2P^2\over 8\pi}={\phi^2_h\over 2\pi}={\Phi^2_h\over 2\pi q^2}~~.
\label{4.3}
\end{equation}
In the case of extreme black hole $\beta^{-1}_H=0$ in (\ref{cond}) and the zero-mode
configuration (\ref{4.1}) is constant $\phi_0=\phi_h$. We find then that $L_0=0$.

Applying now the general formula (\ref{3}) for the entropy of states in a conformal 
field theory we find from (\ref{3.15}) and (\ref{4.3} that)
for a non-extreme black hole
the corresponding entropy
\begin{equation}
S_{conf}=\pi \Phi^2_h=S_{BH}
\label{4.4}
\end{equation}
exactly reproduces the Bekenstein-Hawking expression.
Note that we need only one copy of the Virasoro algebra to get the correct answer.
This is in agreement with the discussion in Section 2.

\section{ Generalization for $d>4$ and $d=3$}
\setcounter{equation}0
The analysis we present above can be extended to dimensions other than four.
The spherically symmetric metric in space-time with $d$ dimensions is
\begin{equation}
ds^2=\gamma_{ab}(x^0,x^1)dx^adx^b+r^2(x^0,x^1)d\Omega^2_{S^{d-2}}~~,
\label{5.1}
\end{equation}
where $d\Omega^2_{S^{d-2}}$ is metric on $(d-2)$-dimensional sphere $S^{d-2}$
of unit radius. General relativity in $d$ dimensions
is described by the action
\begin{equation}
W=-{1\over 16\pi G_d}\int_{M^d}dz^d\sqrt{-g_{(d)}}R_{(d)}~~,
\label{5.2}
\end{equation}
where $G_{(d)}$ is Newton's constant.
Being considered on the class
of metrics (\ref{5.1}) the action (\ref{5.2}) reduces to the effective two-dimensional
theory (omitting the total derivative term)
\begin{equation}
W_{(d)}=-{\Sigma_{d-2}\over 16\pi G_d}\int_{M^2}\left(r^{d-2}R+(d-3)(d-2)r^{d-4}
(\nabla r)^2+(d-3)(d-2)r^{d-4}\right)~~,
\label{5.3}
\end{equation}
where $\Sigma_{d-2}={2\pi^{d-1\over 2}\over \Gamma ({d-1\over 2})}$ is area of the sphere
$S^{d-2}$. Re-defying the dilaton field $r$ as 
\begin{equation}
\Phi^2=Cr^{d-2}~~,~~C={\Sigma_{d-2}\over 2\pi G_d}({d-3\over d-2})~~,
\label{5.4}
\end{equation}
the action (\ref{5.3}) takes the form similar to (\ref{3.2})
\begin{equation}
W_{(d)}=-\int_{M^2}\left({1\over 2}(\nabla\Phi )^2+{1\over 8}({d-2\over d-3})
\Phi^2R+{1\over 8} (d-2)^2C^{2\over d-2}\Phi^{2({d-4\over d-2})}\right)~~.
\label{5.5}
\end{equation}
In $d$ dimensions the horizon is $(d-2)$-dimensional sphere and the Bekenstein-Hawking entropy is
proportional to the area of this sphere
\begin{equation}
S_{BH}^{(d)}={\Sigma_{d-2}\over 4G_d}r_h^{d-2}=({d-2\over d-3}){\pi\over 2}\Phi^2_h~~,
\label{5.6}
\end{equation}
where the value of the  field $\Phi$ on the horizon  is related to the horizon
radius $r_h$ according to (\ref{5.4}).

After  applying the transformation
\begin{equation}
\Phi^2=2({d-3\over d-2})q\Phi_h \phi~~,~~\gamma_{ab}=({\phi_h\over \phi })^{d-3\over d-2}
e^{{2\over q\Phi_h}\phi}\bar{\gamma}_{ab}~~
\label{5.8}
\end{equation}
the action (\ref{5.5}) takes the Liouville type form (\ref{3.3})
\begin{equation}
W_{(d)}=-\int_{M^2}\left({1\over 2}(\nabla\phi )^2+{1\over 4}q\Phi_h \phi R+U_{(d)}(\phi )\right)~~.
\label{5.7}
\end{equation}
As in eq.(\ref{3.3}), $q$ here is an arbitrary parameter.
Note, that the transformation (\ref{5.8}) for the metric is independent of $d$. 
The classical value of the field $\phi$ on the 
horizon is given by
\begin{equation}
\phi_h={1\over 2q}({d-2\over d-3})\Phi_h~~.
\label{5.9}
\end{equation}
Only the potential term $U_{(d)}(\phi )$ in (\ref{5.7}) depends on the dimension $d$.
Its form can be found explicitly but is not important for further consideration. 
As we explained in Section 3, in the region near horizon the potential term effectively disappears
(being multiplied on the function $g(z)$ exponentially decaying at the horizon) in the 
field equation for $\phi$ and the 
constraints. So that
the action (\ref{5.7})
defines at the horizon a conformal theory the form of which is  universal for  any dimension $d$.
 The corresponding Virasoro algebra was analyzed in Sections 3 and 4. It has
\begin{equation}
c=3\pi q^2\Phi^2_h~~~and ~~~~L_0={\phi^2_h\over 2\pi}~~.
\label{5.10}
\end{equation}
From (\ref{5.9}) and (\ref{5.4}) we find that the central charge (\ref{5.10})
\begin{equation}
c=6q^2({d-3\over d-2})S^{(d)}_{BH}
\label{5.11}
\end{equation} 
is proportional to the Bekenstein-Hawking entropy (\ref{5.6}). On the other hand,
we have
\begin{equation}
L_0={1\over 8\pi q^2}({d-2\over d-3})^2\Phi^2_h~~,
\label{5.12}
\end{equation}
where eq.(\ref{5.9}) was used.
Substituting (\ref{5.10})-(\ref{5.12}) 
into Cardy's formula (\ref{3})
\begin{equation}
S_{conf}={\pi\over 2}({d-2\over d-3})\Phi^2_h=S_{BH}^{(d)}
\label{5.13}
\end{equation}
we find
the precise agreement with the Bekenstein-Hawking expression (\ref{5.6})
for the entropy of black hole in $d$ dimensions.

It is seen from the equations (\ref{5.3})-(\ref{5.7}) that the case $d=3$ is special and
should be considered separately. In this case the kinetic and potential terms are absent in (\ref{5.3}).
In order to obtain a non-trivial solution of the gravitational equations we have to add
$\lambda$-term to the Einstein-Hilbert action in three dimensions. The resultant effective 2d theory is
\begin{equation}
W_{(3)}=-{1\over 8G_3}\int_{M^2}\left(r(R+\lambda )\right)~~.
\label{5.14}
\end{equation}
After applying the transformation
\begin{equation}
r=2qG_3\Phi_h\phi~~,~~\gamma_{ab}=e^{{2\over q\Phi_h}\phi}~\bar{\gamma}_{ab}
\label{5.15}
\end{equation}
it takes the form (\ref{5.7}) with the potential
$$
U_{(3)}={1\over  4}\lambda q\Phi_h\phi~ e^{{2\over q\Phi_H}\phi}~~.
$$
Note that in this case $\Phi_h$ is an arbitrary parameter like $q$. We keep it only
in order to illustrate the universality of the action (\ref{5.7}) governing the conformal dynamics
at the horizon  for any $d\geq 3$.  The classical
value of new dilaton field $\phi$ on the horizon now is $\phi_h=(2qG\Phi_h)^{-1}r_h$.
The corresponding conformal field theory  has $c$ and $L_0$ as in (\ref{5.10})
and, as we can see,  the combination $(q\Phi_h)$ indeed drops out in the product $cL_0={3\over 8}{r^2_h\over G^2}$.
Applying the formula (\ref{3}) we then obtain
\begin{equation}
S_{conf}={\pi r_h\over 2 G_3}
\label{5.16}
\end{equation}
in agreement with the Bekenstein-Hawking formula in three dimensions.
Note, that we did not use the Chern-Simons form of the 3d gravity when obtained the correct
answer for the entropy.

\section{  Remarks}
\setcounter{equation}0

\medskip

~~~~~{\bf 6.1 Classical central charge} 

It should be noted that the central charge $c=3\pi q^2\Phi^2_h$ 
of the Virasoro algebra  considered in this paper
is classical. It appears before quantization on the level of the Poisson bracket.
(On the quantum level the central charge is $(1+c)$ and is dominated by the classical value
for large $c$.)
Therefore, it is a reasonable question if namely this value of $c$ describes the degrees of freedom of the theory and should be used in the computation of the entropy by formula (\ref{3}).
This question also arises  \cite{carlip2} in Strominger's calculation of the entropy of
BTZ black hole and is directly related to the problem of degrees of freedom in the
Liouville model (\ref{4}).
Indeed, the large value of the classical central charge $c={3\over 2}{l\over G}$ seems to be in
contradiction with the fact that the Liouville model is a theory of just one scalar field
\cite{SK} and, hence, the effective central charge  $c_{eff}=1$. For Strominger's calculation 
this problem is not yet resolved in
the literature. In our approach, however, there is a hope to overcome this problem.
As we have seen, both the central charge $c$ and $L_0$ (\ref{5.10}) depend on an arbitrary 
parameter $q$ and  do not have an absolute
meaning. Only the combination
$cL_0$, which should be substituted into Cardy's formula, has the absolute
meaning not being dependent on $q$.  Note, that it is an important difference between our approach and the one which   uses the Liouville model where the classical central
charge is fixed and absolute (see, however, \cite{Banados}). 
It should be noted that most of the complications 
of dealing with the Liouville model are due to
the exponential potential term. The conformal field theory (with stress tensor (\ref{3.11}))
appearing in our 
approach is simpler for analysis since the potential
term effectively vanishes at the horizon.

\medskip

{\bf 6.2 Non-spherical gravitational excitations and matter fields} 

In a quantum theory of gravity we should take into account all possible fluctuations of the metric.
Therefore, we should be able to incorporate in our analysis the gravitational excitations which are not
spherical. A way of doing this is to consider all non-spherical excitations as a set of fields propagating on the spherically symmetric background. In this respect they are similar to  the quantum matter
fields and should be considered in the same way. Then, expanding all fields in terms of the 
spherical harmonics $Y_{l,m}$ we obtain an infinite set of fields labeled by $(l,m)$
which are functions on 2d space $M^2$. The corresponding 2d theory can be analyzed and
shown to be conformal at the horizon by the same reasons as in Section 3. The contribution
$S_Q$ of this infinite set of fields to the entropy, though proportional
(in the leading order) to the horizon area, is expected to diverge (either due to the infinite number
of the fields or when one takes the limit of infinite $L$ (see, for example, \cite{F})). 
Presumable, $S_Q$ is what
in the literature  known as the quantum correction \cite{ren} to the Bekenstein-Hawking entropy.
The spherically symmetric excitations of the gravitational field, thus, are responsible for the
``classical'' Bekenstein-Hawking entropy while all other excitations produce a correction.
It was suggested in \cite{ren} that the divergence of $S_Q$
can be absorbed in the renormalization of Newton's constant $G$ so that the total entropy,
$S_{BH}+S_Q$, remains finite. It would be interesting to see how this renormalization 
works in terms of the 2d conformal field theory at the horizon.

\medskip

{\bf 6.3 The Hilbert space and unitary evolution} 

The conformal description we present in this paper may help to
understand the quantum evolution of a system including a black hole as a part.  In this description
one should assign with black hole horizon elements  $|{\cal H}>$ of the Hilbert space realizing a
representation of the Virasoro algebra. In combination with states $| \psi >$ at 
asymptotic infinity they form the complete Hilbert space. 
Considering the evolution in the space of  elements 
$|{\cal H}> \times | \psi > $ there are no reasons why it should not be 
unitary.
Note in this respect the useful analogy between  horizon
and boundary. In a field theory \cite{GZ} on space-time with boundary ${\cal B}$
one should take into account the so-called boundary states $|{\cal B}>$ which live on the boundary.
Only then  an  unitary S-matrix can be constructed.

\bigskip

\bigskip

While this paper was in preparation there appeared an interesting preprint \cite{SC}
which overlaps with our consideration. In particular, both $c$ and $L_0$ 
of the Virasoro algebra found in \cite{SC} 
are proportional to the Bekenstein-Hawking entropy
though   depend on the choice of period. Only the combination $cL_0$ is unambiguous.
This is in agreement with our result.

I thank Steve Carlip for correspondence and suggestion to generalize my analysis to
higher dimensions. I also would like to thank Gerard 't Hooft for interesting discussions.

\bigskip

{\it Note added:} It is interesting to note that in the Section 2 the equation 
$x_-=0$ determining the location of the horizon $\cal H$ can be replaced by a more general
one: $x_-=f(x_+)$, $(f'(x_+)\neq 0)$. The horizon in this case is dynamically evolving.
Introducing then new coordinates $z_+=x_-+f(x_+)$ and
$z_-=x_- -f(x_+)$ the horizon is now located at $z_-=0$. The condition (2.2) can be 
re-formulated near $z_-=0$ as $(\nabla r)^2=\lambda (z_+)z_-+O(z^2_-)$. 
It is invariant under diffeomorphisms
generated by vector $\hat{\xi}=\xi (z_+)\partial_{z_+}$. The basis vectors
$\hat{\xi}_n=e^{\imath nz_+}\partial_{z_+}$ again form the Virasoro algebra.

\end{document}